\documentclass[preprint,prd,nofootinbib,tightenlines,amsmath]{revtex4}
\usepackage{axodraw}
\usepackage{graphics}
\usepackage{epsfig}
\usepackage{dcolumn}
\usepackage{bm}

\begin{document}
\baselineskip=15pt
\parskip=5pt

\hspace*{\fill} $\hphantom{-}$

\def\lsim{\mathrel {\vcenter {\baselineskip 0pt \kern 0pt
    \hbox{$<$} \kern 0pt \hbox{$\sim$} }}}
\def\gsim{\mathrel {\vcenter {\baselineskip 0pt \kern 0pt
    \hbox{$>$} \kern 0pt \hbox{$\sim$} }}}

\preprint{hep-ph/0610362}

\title{Has HyperCP Observed a Light Higgs Boson?}

\author{Xiao-Gang He}
\email{hexg@phys.ntu.edu.tw} \affiliation{Department of Physics
and Center for Theoretical Sciences, National Taiwan University, Taipei}

\author{Jusak Tandean}
\email{jtandean@ulv.edu}
\affiliation{Department of Mathematics/Physics/Computer Science,
University of La Verne, La Verne, CA 91750, USA}

\author{G. Valencia}
\email{valencia@iastate.edu}
\affiliation{Department of Physics and Astronomy, Iowa State University, Ames, IA 50011, USA}

\date{\today}

\begin{abstract}

The HyperCP collaboration has observed three events for the decay
$\Sigma^+\to p\mu^+\mu^-$ which may be interpreted as a new particle of mass 214.3\,MeV.
However, existing data from kaon and $B$-meson decays severely constrain  this interpretation,
and it is nontrivial to construct a model consistent with all the data.
In this letter we show that the ``HyperCP particle'' can be identified with the light
pseudoscalar Higgs boson in the next-to-minimal supersymmetric standard model, the $A_1^0$.
In this model there are regions of parameter space where the $A_1^0$ can satisfy all the
existing constraints from kaon and $B$-meson decays and mediate
$\Sigma^{+}\to p \mu^{+}\mu^{-}$  at a level consistent with the HyperCP observation.

\end{abstract}


\maketitle

Three events for the decay mode  \,$\Sigma^+\to p\mu^+\mu^-$\,  with a dimuon invariant
mass of 214.3\,MeV have been recently observed by the HyperCP Collaboration~\cite{Park:2005ek}.
It is possible to account for these events within the Standard Model (SM) when long-distance
contributions are properly included~\cite{He:2005yn}.
However, the probability of having all three events at the same dimuon mass,
given the SM predictions,  is less than one percent.
This suggests a new-particle interpretation for these events, for which the branching ratio
is  \,$\bigl(3.1^{+2.4}_{-1.9}\pm1.5\bigr)\times 10^{-8}$~\cite{Park:2005ek}.\,

This possibility has been explored to some extent in the
literature~\cite{He:2005we,Deshpande:2005mb,Geng:2005ra}, where it has been shown
that kaon decays place severe constraints on the flavor-changing two-quark couplings
of the hypothetical new particle, $X$.
It has also been claimed that a light sgoldstino is a viable
candidate~\cite{Gorbunov:2000cz}.
It is well known in the case of light Higgs production in kaon decay that, in addition to
the two-quark flavor-changing couplings, there are comparable four-quark contributions~\cite{sdH}.
They arise from the combined effects of the usual SM four-quark  \,$|\Delta S|=1$\, operators
and the flavor-conserving couplings of $X$.
We have recently computed the analogous four-quark contributions to light Higgs production in
hyperon decay~\cite{long} and found that they can also be comparable to
the two-quark contributions previously discussed in the literature.

The interplay between the two- and four-quark contributions makes it possible to find models
with a light Higgs boson responsible for the HyperCP events that has not been observed in
kaon or $B$-meson decay.
However, it is not easy to devise such models respecting all the experimental constraints.
In most models that can generate  $\bar d s X$  couplings, the two-quark operators have
the structure  \,$\bar d(1\pm\gamma_5^{})s X$.\,
Since the part without $\gamma_5^{}$ contributes significantly to  \,$K\to\pi\mu^+\mu^-$,\,
their data imply that these couplings are too small to account for the HyperCP
events~\cite{He:2005we,Deshpande:2005mb,Geng:2005ra}.
In some models, there may be parameter space where the four-quark contributions mentioned
above and the two-quark ones are comparable and cancel sufficiently to lead to suppressed
\,$K\to\pi\mu^+\mu^-$\,  rates while yielding \,$\Sigma^+\to p\mu^+\mu^-$\, rates
within the required bounds.
However, since in many models the flavor-changing two-quark couplings  $\bar q q'X$ are
related for different $(q,q')$ sets, experimental data on $B$-meson decays,
in particular  \,$B\to X_s\mu^+\mu^-$,\,  also provide stringent constraints.
For these reasons, the light (pseudo)scalars in many well-known models, such as the SM
and the two-Higgs-doublet model, are ruled out as candidates to explain the HyperCP
events~\cite{long}.

In this paper we show that a light pseudoscalar Higgs boson,
$A_1^0$,  in the next-to-minimal supersymmetric standard model
(NMSSM)~\cite{nmssmodel} can be identified with the $X$ particle
satisfying the constraints
\begin{eqnarray}
{\cal B}(K^\pm \to \pi^\pm A_1^0) \,\,\lesssim\,\, 8.7 \times 10^{-9} \,\,, &&
{\cal B}(K_S^{} \to \pi^0 A_1^0) \,\,\lesssim\,\, 1.8 \times 10^{-9} \,\,, \nonumber \\
{\cal B}(B\to X_s^{} A_1^0) &\lesssim& 8.0 \times 10^{-7} \,\,,  \vphantom{\sum^x}
\label{bounds}
\end{eqnarray}
obtained in Ref.~\cite{long} from  the measurements of
\,$K^\pm\to\pi^\pm\mu^+\mu^-$~\cite{Ma:1999uj,Park:2001cv},
\,$K_S\to\pi^0\mu^+\mu^-$~\cite{Batley:2004wg}, and
\,$B\to X_s\mu^+\mu^-$~\cite{Aubert:2004it,Iwasaki:2005sy}.
We include both two- and four-quark contributions to the $K$ and $\Sigma^+$ decays,
neglecting the latter contributions to the $B$ decay.

In the NMSSM there is a gauge-singlet Higgs field $N$ in addition to the two Higgs fields
$H_u$ and $H_d$ responsible for the  up- and down-type quark masses in the MSSM.
We follow the specific model described in Ref.~\cite{Hiller:2004ii}, with suitable modifications.
The superpotential of the model is given by
\begin{eqnarray}
W &=& Q Y_u^{} H_u^{} U + Q Y_d^{} H_d^{} D + L Y_e^{} H_d^{} E + \lambda H_d^{} H_u^{} N
- \mbox{$\frac{1}{3}$} k N^3  \,\,,
\end{eqnarray}
where  $Q$, $U$, $D$, $L$, and $E$ represent the usual quark and lepton fields, $Y_{u,d,e}$
are the Yukawa couplings, and $\lambda$ and $k$ are dimensionless parameters.
The soft-supersymmetry-breaking term in the Higgs potential is
\begin{eqnarray}
V_{\rm soft}^{} &=& m^2_{H_u} |H_u^{}|^2 + m^2_{H_d} |H_d^{}|^2 + m^2_N |N|^2
- \left(\lambda A_\lambda^{}H_d^{}H_u^{}N+\mbox{$\frac{1}{3}$}k A_k^{}N^3+{\rm H.c.}\right)  \,\,,
\end{eqnarray}
and the resulting Higgs potential  has a global U(1)$_R$ symmetry in the limit that the
parameters  \,$A_\lambda,A_k\to0$\,~\cite{Dobrescu:2000yn}.

There are two physical pseudoscalar bosons in the model which are linear combinations of
the pseudoscalar components in $H_{u,d}$ and~$N$.
The lighter one, $A^0_1$, has a mass given  by
\begin{eqnarray}
m^2_{\cal A} &=& 3 k\, x\, A_k^{}  \,+\,  {\cal O}(1/\tan\beta)
\end{eqnarray}
in the large-$\tan\beta$ limit, where  \,$x=\langle N\rangle$\, is
the vacuum expectation value of $N$  and  $\tan\beta$ is the ratio
of vacuum expectation values of the two Higgs doublets. If the
U(1)$_R$ symmetry is broken slightly, the mass of $A^0_1$ becomes
naturally small, with values as low as  \,$\sim$100\,MeV\,
phenomenologically allowed~\cite{Hiller:2004ii,Dobrescu:2000yn}.
We will now show that this $CP$-odd scalar can play the role of
the $X$ particle.

At tree level, the $A^0_1$ couplings to up- and down-type quarks and to leptons in this model
are diagonal and described by
\begin{eqnarray}
{\cal L}_{{\cal A}q q}^{}  \,\,=\,\,
-\left(l_u^{}m_u^{}\,\bar u\gamma_5^{}u + l_d^{}m_d^{}\,\bar d\gamma_5^{}d\right)
\frac{i A^0_1}{v}  \,\,,  \hspace{2em}
{\cal L}_{{\cal A}\ell}^{}  \,\,=\,\,
\frac{i g_\ell^{} m_\ell^{}}{v}\, \bar\ell\gamma_5^{}\ell\, A^0_1
\label{44q}
\end{eqnarray}
in the general notation of Ref.~\cite{long} used to compute the effect of four-quark operators,
where
\begin{eqnarray}
l_d^{} \,\,=\,\, -g_\ell^{} \,\,=\,\, \frac{v\,\delta_-^{}}{\sqrt{2}\,x}  \,\,, \hspace{2em}
l_u^{} \,\,=\,\, \frac{l_d}{\tan^2\beta}   \,\,.
\end{eqnarray}
with  \,$v=246$\,GeV\, being the electroweak scale and
\,$\delta_-^{}=(A_\lambda-2k x)/(A_\lambda+k x)$.\,
Thus,  $l_u$ can be neglected in the large-$\tan\beta$ limit.
The interactions in ${\cal L}_{{\cal A}q q}$ combined with the operators due to
$W$ exchange between quarks induce the four-quark contributions discussed earlier.

For an $A^0_1$ of mass 214.3\,MeV as we propose, the decay into
a muon-antimuon pair completely dominates over the other kinematically
allowed modes:  $A^0_1 \to e^+e^-,\, \nu\bar{\nu},\, \gamma\gamma$.\,
Accordingly, we assume  \,${\cal B}(X\to\mu^+\mu^-)\sim 1$.\,
In addition, the muon anomalous magnetic moment imposes the constraint~\cite{He:2005we}
\begin{equation}
|g_\ell^{}| \,\,\lesssim\,\, 1.2  \,\,.  \label{gmu}
\end{equation}
This bound results in an $A^0_1$ width  \,$\Gamma_{\cal A}\lesssim3.7\times10^{-7}$\,MeV,\,
which is consistent with the estimate of $10^{-7}$\,MeV for the width of the HyperCP
particle~\cite{Geng:2005ra} if  \,$|g_\ell|=|l_d|={\cal O}(1)$.\,
An $l_d$ of order one is also consistent with the general estimates~\cite{Hiller:2004ii}
for the size of  \,$v\delta_-^{}/\bigl(\sqrt2\,x\bigr)$.

At one-loop level, the two-quark flavor-changing Lagrangian for  \,$s\to d A^0_1$\,  can be
written as
\begin{eqnarray} \label{quark}
{\cal L}_{{\cal A}sd}^{} &=& \frac{ig_{\cal A}^{}}{v} \left[
m_s^{}\,\bar{d}(1+\gamma_5^{})s \,-\,
m_d^{}\,\bar{d}(1-\gamma_5^{})s \right] A_1^0 \,\,+\,\,  {\rm
H.c.}  \,\,, \label{penguinlag}
\end{eqnarray}
where
\begin{eqnarray}
g_{\cal A}^{} &=&
\frac{G_F^{}}{8\sqrt2\,\pi^2}\sum_{q=u,c,t}V^*_{qd}V_{qs}^{}\,C_{{\cal A}q}^{} \,\,,
\label{penguin}
\end{eqnarray}
with  $V_{qr}$ being the elements of the Cabibbo-Kobayashi-Maskawa (CKM) matrix and
$C_{{\cal A}q}^{}$  depending on the particles in the loops~\cite{Hiller:2004ii}.
The corresponding coupling constant $g_{\cal A}^\prime$ in \,$b\to s A^0_1$\,
has a similar expression in terms of  \,$V^*_{qs}V_{qb}^{}\,C_{{\cal A}q}^{}.$\,

In the large-$\tan\beta$ limit, the pseudoscalar $A^0_1$ in this model does not couple to
up-type quarks and  Eq.~(\ref{penguin}) is completely determined by parameters in
the supersymmetry (SUSY) sector.
In particular, $C_{{\cal A}q}^{}$ can be zero if a super-Glashow-Iliopoulos-Maiani
mechanism is operative~\cite{Hiller:2004ii}.
In this limit, the decays  \,$\Sigma^+\to p A^0_1$\, and \,$K\to\pi A^0_1$\,  are dominated by
four-quark contributions~\cite{long}, and the mode  \,$b\to s A^0_1$\,  is negligible.
However, the results of Ref.~\cite{long} imply that a vanishing  $g_{\cal A}^{}$  due to
\,$C_{{\cal A}q}^{}\to0$\, would not satisfy all the existing constraints.

This leads us to some details of the model. Assuming that there is
only stop mixing, and that sup and scharm are degenerate, we have
for  \,$s\to d A^0_1$\, or  \,$b\to s A^0_1$~\cite{Hiller:2004ii},
\begin{eqnarray}
C^{}_{{\cal A}q} &=& \delta_{qt}\,\tan\beta \left[ -l_d^{} \sum_{i=1,2} X_i
+ \lambda\, m_t^{} v\, \sin\theta_{\tilde t}^{}\,\cos\theta_{\tilde t}^{}
\sum_{u,w=1,2}Y_{uw}\right]  \,\,,
\label{hilor}
\end{eqnarray}
where $X_i$ and $Y_{uw}$ are loop-integral functions of parameters in the chargino and squark
mixing matrices given in Ref.~\cite{Hiller:2004ii}, and  $\theta_{\tilde t}$ is the stop
mixing angle.
Consequently, in this scenario the CKM factors involving the top-quark are the relevant ones.
It then follows from Eq.~(\ref{penguin}) that  $C^{}_{{\cal A}q}$  of the right magnitude
for  \,${\cal B}(\Sigma^+\to p A_1^0)\sim3\times10^{-8}$\,  will yield
\,${\cal B}(B\to X_s A_1^0)$\, that violates its limit in Eq.~(\ref{bounds}) by
about three orders of magnitude.

Now, the penguin amplitude in this model is dominated by squark and chargino intermediate
states, and the particular scaling with the top-quark CKM factor in Eq.~(\ref{hilor})
follows from the assumption of stop mixing only.
If we assume instead that there is only sup mixing,  $C^{}_{{\cal A}q}$  will be
proportional to $\delta_{qu}$.
In this case, it is possible for $A_1^0$ to be the HyperCP particle without violating
the bounds in Eq.~(\ref{bounds}).
To be specific, assuming that sup mixing is dominant, and that stop and scharm
are degenerate, we obtain
\begin{eqnarray}
C^{}_{{\cal A}q} &=& \delta_{qu} \tan\beta \left[-l_d^{} \sum_{i=1,2} X_i^{}
+ \lambda\, m_u^{}v\, \sin\theta_{\tilde u}^{}\,\cos\theta_{\tilde u}^{}
\sum_{u,w=1,2}Y_{uw}\right]  \,\,.
\label{hilprime}
\end{eqnarray}
In addition,  Eq.~(\ref{hilprime}) has a vanishing $CP$-violating phase
if all the SUSY parameters are real.
This is needed if the contributions from two- and four-quark operators are to partially
cancel each other~\cite{long}, as needed to satisfy the kaon bounds in Eq.~(\ref{bounds}).
To simplify the discussion, we now consider the case where  \,$\theta_{\tilde u}=0$.\,
In this case
\begin{eqnarray}  \label{gA}
g_{\cal A}^{} &=& \frac{G_F}{8\sqrt2\,\pi^2}\, V_{ud}^*V_{us}^{}\, \sqrt2\, m_W^{}\, \tan\beta
\sum_i m_{\chi_i}^{}U_{i2}^{}V_{i1}^{} \left[ D_3^{}\bigl(y_{c,i}^{}\bigr)
- D_3^{}\bigl(y_{u,i}^{}\bigr) \right] l_d^{}  \,\,,
\end{eqnarray}
where $U_{ij}$ and $V_{ij}$ are the unitary matrices diagonalizing the chargino mass matrix,
\,$D_3(y)=y\,\ln y/(1-y)$\,  is a loop-integral function, and
\,$y_{r,i}^{}=m^2_{\tilde r}/m^2_{\chi_i}$,\,  with  $m_{\chi_i}$  being a chargino
mass~\cite{Hiller:2004ii}.
Therefore, depending on whether the sup mass is larger or smaller than the scharm mass,
$g_{\cal A}^{}$ can take either sign.

The expressions for the four-quark contributions to the kaon and hyperon decays are lengthy and
in general depend on the effective coupling of $A_1^0$ to a gluon pair~\cite{long}.
For the simple case above, the contributions to $A^0_1 gg$ are from quarks in the loop in
the large-$\tan\beta$ limit, and the squarks do not contribute since there is no left- and
right-squark mixing.
Since \,$l_u\to 0$,\, the four-quark contributions can be written purely in terms of $l_d$.
Using the results in Ref.~\cite{long} (with \,$D-F=0.25$\,),  the full amplitudes are
\begin{eqnarray}
i{\cal M}(K^\pm\to\pi^\pm A_1^0) &=& g_{\cal A}^{}\,\frac{m_K^2-m_\pi^2}{v}
- \left(1.08 \times 10^{-7}\right) l_d^{}\, \frac{m_K^2}{v}  \,\,,
\end{eqnarray}
\begin{eqnarray}
{\cal M}(\Sigma^+\to p A_1^0) &=& \left[
g_{\cal A}^{}\,\frac{m_\Sigma^{}-m_p^{}}{v}
- \left(6.96\times 10^{-7}\right)l_d^{}\, \frac{f_\pi^{}}{v}\right] i \bar{p}\Sigma^+
\nonumber \\ && \!\! - \,\,
\left[(0.25)\, g_{\cal A}^{}\,\frac{m_\Sigma^{}+m_p^{}}{v}\,\frac{m_K^2}{m_K^2-m_{\cal A}^2}
+ \left(5.72\times 10^{-6}\right)l_d^{}\, \frac{f_\pi^{}}{v} \right]
i \bar{p}\gamma_5^{}\Sigma^+.
\end{eqnarray}
The corresponding amplitude for  \,$K_S\to\pi^0A_1^0$,\,  which we have not displayed,
can be derived from Ref.~\cite{long}.

\begin{figure}[b] \vspace{4ex}
\includegraphics[width=4.5in]{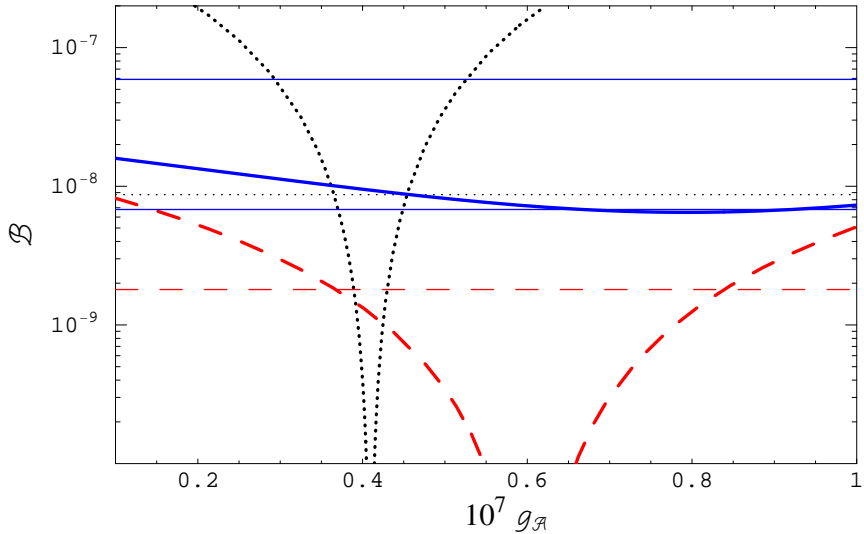}   \vspace{-1ex}
\caption{\label{figbr}%
Branching ratios of  \,$\Sigma^+\to p A_1^0$\, (solid curve), \,$K^+\to\pi^+A_1^0$\,
(dashed curve), and  \,$K_S\to\pi^0A_1^0$\,  (dotted curve)
as  functions of $g_{\cal A}^{}$  for  \,$l_d=0.35$.\,
The horizontal lines indicate the bounds from the HyperCP result and
the kaon bounds from  Eq.~(\ref{bounds}).
The allowed range in this case is  $\,0.38\lesssim10^7\,g_{\cal A}^{}\lesssim0.43.\,$
}\end{figure}

Choosing, for example, \,$l_d=0.35$,\, we show the resulting branching ratios in Fig.~\ref{figbr}.
Even though it is possible for $A_1^0$ to explain the HyperCP observation for a wide range of
parameters, as the figure indicates, we have found that there are only  rather narrow ranges
of $g_{\cal A}^{}$  and  $l_d$ for which the kaon bounds are also satisfied, namely
\,$1.0\times10^{-7}\lesssim g_{\cal A}^{}/l_d\lesssim 1.3\times10^{-7}$.\,
The $B$-meson bound is automatically satisfied with our assumptions above.

We remark that $g_{\cal A}^{}$ in Eq.~(\ref{gA}) is proportional
to \,$G_F V_{ud}^*V_{us}^{}\,\tan\beta\, m_W^{} m_{\chi_i}$,\,  which is of $\cal O$(1) for
\,$\tan\beta=\cal O$(30)\, and  \,$m_{\chi_i}=\cal O$(100\,GeV).\,
Thus, in order to have  \,$g_{\cal A}^{}={\cal O}(10^{-7})$\, with these parameter
choices, one needs
\,$U_{i2}V_{i1}\,\bigl[D_3^{}\bigl(y_{c,i}^{}\bigr)-D_3^{}\bigl(y_{u,i}^{}\bigr)\bigr]$\,
to be of order  $10^{-7}$  as well.
Since as mentioned above there is a super-GIM mechanism that can be at work, as long as
$m_{\tilde u}$ and $m_{\tilde c}$ are degenerate enough, one can get an arbitrarily small
$g_{\cal A}^{}$ to satisfy the required value.
It turns out that there is also  SUSY parameter space where the desired value of
$g_{\cal A}^{}$ can be produced without very fine-tuned degeneracy.

\begin{figure}[b] \vspace{2ex}
\includegraphics[width=3in]{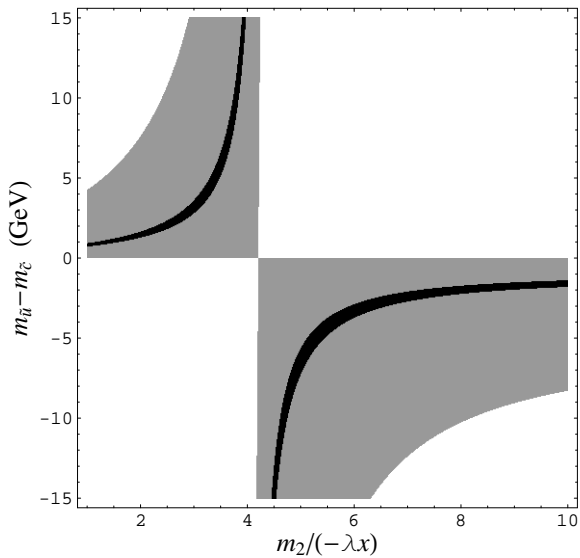}   \vspace{-1ex}
\caption{\label{susy}%
Parameter space for  \,$m_{\tilde u}$$-$$m_{\tilde c}$\,  and  $m_2^{}/(-\lambda x)$  where
$A_1^0$ can explain the HyperCP events (gray regions) and simultaneously satisfy the kaon
bounds (black regions), with the input described in the text.}
\end{figure}

We have carried out a detailed exploration of the parameter space.
In general, if the chargino masses are larger than the squark masses, fine-tuned squark
masses, with splitting of less than one GeV between sup and scharm masses, are required.
However, if the chargino masses  [of ${\cal O}$(100\,GeV)] are smaller than
the squark masses [of ${\cal O}$(1\,TeV)], there is some parameter space satisfying
the constraints with squark-mass splitting of a few GeV to tens of GeV.
Taking  \,$-\lambda x=150$\,GeV,\,  \,$m_{\tilde c}=2.5$\,TeV,\,  and  \,$\tan\beta =30$,\,
we display in Fig.~\ref{susy} the allowed ranges of  \,$m_{\tilde u}$$-$$m_{\tilde c}$\,
and  \,$m_2^{}/(-\lambda x)$\,  corresponding to the required values
\,$1.0\times10^{-7}\lesssim g_{\cal A}^{}/l_d\lesssim 1.3\times10^{-7}$,  where  $m_2^{}$ is
the first element of the chargino undiagonalized mass-matrix~\cite{Hiller:2004ii}.
In this case, the two chargino eigenmasses  are, respectively,
\,$100{\rm\,GeV}\lesssim m_{\chi_1}\lesssim150{\rm\,GeV}$\,  and
\,$200{\rm\,GeV}\lesssim m_{\chi_2}\lesssim1500{\rm\,GeV}$.\,
These chargino masses, especially the lighter one, are in the
interesting range which may be probed at the LHC and ILC.
In this figure, we also display the much larger (gray) regions where $A_1^0$ can explain
the HyperCP observation.

In conclusion, we have shown that a light pseudoscalar Higgs in the NMSSM can be identified
as the possible new particle responsible for the HyperCP events while satisfying all
constraints from kaon and $B$-meson decays.
We have found that there are large regions in the parameter space where the particle can
explain the HyperCP observation.
This by itself is not trivial, considering that the SM, for example, yields rates that
are too large by more than a factor of ten~\cite{long}.
Only after the kaon bounds are imposed does the allowed parameter space become much more
restricted.
We believe that this is sufficiently intriguing to warrant a revisiting of these kaon bounds,
perhaps by the NA48 experiment.

\begin{acknowledgments}

The work of X.G.H. was supported in part by NSC and NCTS. The work of G.V. was supported
in part by DOE under contract number DE-FG02-01ER41155.

\end{acknowledgments}

\end{document}